\documentclass{article}
\usepackage{graphicx}
\usepackage{amsmath}
\usepackage{amsfonts}
\usepackage{amssymb}

\begin{document}

\title{Localization of gravitational energy in ENU model and its consequences}
\author{Jozef Sima, Miroslav Sukenik and Julius Vanko}
\maketitle
\begin{abstract}
The contribution provides the starting points and background of the model of
Expansive Nondecelerative Universe (ENU), manifests the advantage of
exploitation of Vaidya metrics for the localization and quantization of
gravitational energy, and offers four examples of application of the ENU
model, namely energy of cosmic background radiation, energy of Z and W bosons
acting in weak interactions, hyperfine splitting observed for hydrogen 1s
orbital. Moreover, time evolution of vacuum permitivity and permeability is predicted.
\end{abstract}

\section{Section}

$\backslash$%
$\backslash$%

Localization of gravitational energy in ENU model and its consequences

Jozef Sima$^{a}$, Miroslav Sukenik$^{a}$ and Julius Vanko$^{b}$

$^{a}$Slovak Technical University, Dep. Inorg. Chem., Radlinskeho 9, 812 37
Bratislava, Slovakia

$^{b}$Comenius University, Dep. Nucl. Physics, Mlynska dolina F1, 842 48
Bratislava, Slovakia

e-mail: sima@chelin.chtf.stuba.sk; vanko@fmph.uniba.sk

Abstract. The contribution provides the starting points and background of the
model of Expansive Nondecelerative Universe (ENU), manifests the advantage of
exploitation of Vaidya metrics for the localization and quantization of
gravitational energy, and offers four examples of application of the ENU
model, namely energy of cosmic background radiation, energy of Z and W bosons
acting in weak interactions, hyperfine splitting observed for hydrogen 1s
orbital. Moreover, time evolution of vacuum permitivity and permeability is predicted.

I. Theoretical backround

Due to the simultaneous creation of both the matter and gravitational energy
(having the identical absolute values but differing in the sign of the values)
the total energy of the Universe is equal to zero in the model of Expansive
Nondecelerative Universe (ENU) and thus one of the fundamental requirement of
the Universe evolution [1] is fulfilled. It has been evidenced [2] that such a
Universe can expand by the velocity of light $c$ and it therefore holds

$a=c.t_{c}=\frac{2G.M_{U}}{c^{2}}\qquad\qquad\qquad$ (1)

where $a$ is the gauge factor (at present a 1.3 x 10$^{26}$ m), $t_{c}$is the
cosmological time, $M_{U}$ is the mass of the Universe (it approaches at
present 8.6 x 10$^{52}$ kg).

In the ENU model, due to the matter creation the Vaidya metrics [3] must be
used which enables to localize the gravitational energy. For weak fields
Tolman's relation [4]

$\varepsilon_{g}=-\frac{R.c^{4}}{8\pi.G}=-\frac{3m.c^{2}}{4\pi a.r^{2}}$
\qquad\qquad(2)

can be applied in which $\varepsilon_{g}$ is the density of the gravitational
energy induced by a body with the mass $m$ in the distance $r,R$ denotes the
scalar curvature. It should be pointed out that contrary to the more
frequently used Schwarzschild metrics (in which $\varepsilon_{g}$= 0 outside a
body, and $R$ = 0), in the Vaidya metrics $R$ $\neq$0 and $\varepsilon_{g}$
may thus be quantified and localized also outside a body. It has been shown
[4] that at the same time it must hold

$\varepsilon_{g}=\frac{3E_{g}}{4\pi.\lambda^{3}}$ \qquad\qquad\qquad\qquad\ \ \ (3)

where $E_{g}$ is the quantum of the gravitational energy, the corresponding
Compton wavelength can be expressed as

$\lambda=\frac{\hbar.c}{E_{g}}$ \qquad\qquad\qquad\qquad\ \ \ \ \ \ \ (4)

Substitution of (4) into (3) and comparison of (2) and (3) leads to

$\left|  E_{g}\right|  =\left(  \frac{m.\hbar^{3}.c^{5}}{a.r^{2}}\right)
^{1/4}$ \qquad\qquad\ \ \ (5)

in which $E_{g}$denotes the quantum of the gravitational energy induced by a
body with the mass $m$in the distance $r$.

The validity of (5) was tested both in the field of macrosystems and
microworld. Application of equation (5) allowed us to derive in an independent
way the Hawking's relation for black hole evaporation and explain the presence
of some peaks in low-temperature far-infrared and Raman spectra of several
compounds [4].

Some of the further verifications and applications of relation (5) are given
in the following parts.

II. Energy of cosmic background radiation

From the beginning to the end of radiation era, the Universe was in
thermodynamic equilibrium. Based on the above postulate it can be supposed
that the energy of a photon of the cosmic background radiation equaled to the
energy of a gravitational quantum, i.e.

$k.T=\left|  E_{g}\right|  =\left(  \frac{m.\hbar^{3}.c^{5}}{a.r^{2}}\right)
^{1/4}$ \qquad(6)

When taking m in (6) as the mass of the Universe, $M_{U}$

$M_{U}=\frac{a.c^{2}}{2G}$ \qquad\qquad\qquad\qquad\ \ \ \ (7)

and $r$ as the gauge factor $a$

$r=a$ \qquad\qquad\qquad\qquad\qquad\ \ \ \ \ \ (8)

a well-known formula [5]

$k.T\cong\left(  \frac{\hbar^{3}.c^{5}}{2G.t_{c}^{2}}\right)  ^{1/4}$
\qquad\qquad\qquad\ (9)

is obtained, however, when comparing to [5], the mode of its derivation is
independent. The present consistency might be evaluated as an evidence of
justification of the ENU model.

III. Weak interactions

In our previous paper [6] the mass of Z and W bosons was derived stemming from
the energy density. As it will be shown in the following, an identical
relationship can be obtained using equation (5), i.e. stemming from
gravitational energy quantization. Let us substitute m by the limiting mass [6]

$m=\frac{a.\hbar^{2}}{g_{F}}$ \qquad\qquad\qquad\qquad\ \ \ \ \ \ (10)

where $g_{F}$ is the Fermi constant, and express $r$ as the Comptom wavelength
of the vector bosons Z and W possessing the mass $m_{ZW}$

$r=\frac{\hbar}{m_{ZW}.c}$ \qquad\qquad\qquad\qquad\ \ \ \ \ (11)

In such a case, from (5), (10) and (11) we obtain

$\left|  E_{g}\right|  =m_{ZW}.c^{2}$ \qquad\qquad\qquad\ \ \ \ (12)

if the known relation

$m_{ZW}^{2}\cong\frac{\hbar^{3}}{g_{F}.c}$ \qquad\qquad\qquad\qquad\ \ (13)

was applied.

IV. Hyperfine structure of the hydrogen atom K-level

Equation (5) can be exploited to an independent prediction of the value of
hyperfine structure $E_{HF}$observed in the spectra of hydrogen atom
(experimental value for the electron occupying H1s orbital is $E_{HF}$ = 1420
MHz). Suppose, the energy of the hyperfine splitting induced in the hydrogen
atom K-level by the proton magnetic momentum is identical to the energy given
by equation (5). Such an identity may be taken as a condition of the stability
of the hydrogen atom. When putting the mass of electron $m_{e}$ (9.109 x
10$^{-31}$ kg) and the Bohr radius of H1s orbital

$r\cong52.9\times10^{-12}m$ \qquad\qquad\ \ \ \ (14)

into (5), the energy value

$E_{HF}\cong2400MHz$ \qquad\qquad\qquad(15)

is obtained. This value is 1.7 times higher than the experimental value and
thus closer to it than that of calculated one using a commonly applied
simplified equation (16)

$E_{HF}\cong\frac{I_{H1s}.\alpha^{2}.m_{e}}{m_{p}}$ \qquad\qquad\qquad\ (16)

in which $I_{H1s}$ is the hydrogen atom ionization energy (13.6 eV) and
$\alpha$is the constant of hyperfine splitting.

V. Time evolution of the vacuum permitivity

The constant of hyperfine structure $\alpha$ is defined as

$\alpha=\frac{e^{2}}{4\pi.\epsilon_{o}.\hbar.c}$ \qquad\qquad\qquad\qquad(17)

At the beginning of separation of electromagnetic interactions the equation

$\alpha=1$ \qquad\qquad\qquad\qquad\qquad\ \ \ (18)

had to be valid. When substituting

$r=\frac{\hbar}{m_{e}.c.\alpha}$ \qquad\qquad\qquad\qquad\ \ \ (19)

into the left side of (21) and

$I_{H1s}\cong m_{e}.c^{2}.\alpha^{2}$ \qquad\qquad\qquad(20)

into the right side of\ (21)

$\left(  \frac{m_{e}.\hbar^{3}.c^{5}}{a.r^{2}}\right)  ^{1/4}\cong
I_{H1s}.\alpha^{2}.\frac{m_{e}}{m_{p}}$ \ \ \ \ (21)

the dependence

$\alpha\approx a^{-1/14}$ \qquad\qquad\qquad\qquad\ \ \ (22)

appears. Two conclusions may be derived from the above relationships. The
first one is that equation (18) relates to the time

$t\cong10^{-10}s$ \qquad\qquad\qquad\qquad\ \ \ (23)

which is just the time in which the weak and electromagnetic interactions were
separated. The second consequence relates to (22), i.e. the time evolution of
the constant of hyperfine splitting. Since the velocity of light, electronic
charge and Planck constant are considered to be time independent quantities,
time evolution of the Universe (e.g. changes in its mass and, in turn, also in
charge and electrostatic field density) and the gradual increase of the gauge
factor may be reflected in a very slow change in the vacuum permitivity
$\varepsilon_{o}$ (an electric property) and vacuum permeability $\mu_{o}$ (a
magnetic property).

Conclusions

1.Increase in the gauge factor has several consequences which are to be
unveiled and explained in the future. One of them is a gradual decrease in the
hyperfine splitting constant which can be related to a time-increasing of
vacuum permitivity and time-decreasing of vacuum permeability.

2.Capability of localization of the gravitational energy within ENU is a
challenge for answering the questions such as unification of all four
fundamental physical interactions, stability or invariability of some physical
quantities and ``constants``.

References

1.S. Hawking, Sci. Amer., 236 (1980) 34

2.V. Skalsky, M. Sukenik, Astrophys. Space Sci., 236 (1991) 169

3.P.C. Vaidya: Proc. Indian Acad. Sci., A33 (1951) 264

4.J. Sima, M. Sukenik, Preprint: gr-qc 9903090

5. I. L. Rozentahl, Advances in Mathematics, Physics and Astronomy, 31 (1986)
241 (in Czech)

6.M. Sukenik, J. Sima, J. Vanko, Preprint: gr-qc 0001059
\end{document}